\documentclass[sigconf]{acmart}

\usepackage{times}
\usepackage{soul}
\usepackage{url}
\usepackage[utf8]{inputenc}
\usepackage{graphicx}
\usepackage{amsmath}
\usepackage{amsthm}
\usepackage{booktabs}
\usepackage{subfig}
\usepackage{amsfonts}
\usepackage{multirow}
\usepackage{algorithm}
\usepackage{algorithmic}
\usepackage{booktabs}
\usepackage{amsmath}
\usepackage{caption}
\usepackage{subfig}
\usepackage{xspace}
\usepackage{bbm}
\usepackage{textcomp}

\newcommand{\dglke}{DGL-KE\xspace}

\DeclareMathOperator{\diag}{diag}

\AtBeginDocument{%
  \providecommand\BibTeX{{%
    \normalfont B\kern-0.5em{\scshape i\kern-0.25em b}\kern-0.8em\TeX}}}

\begin{document}

\title{DGL-KE: Training Knowledge Graph Embeddings at Scale}

\author{Da Zheng}
\email{dzzhen@amazon.com}
\affiliation{%
  \institution{AWS AI}
}

\author{Xiang Song}
\email{xiangsx@amazon.com}
\affiliation{%
  \institution{AWS Shanghai AI Lab}
}

\author{Chao Ma}
\email{manchao@amazon.com}
\affiliation{%
  \institution{AWS Shanghai AI Lab}
}

\author{Zeyuan Tan}
\email{zeyut@amazon.com}
\affiliation{%
  \institution{AWS Shanghai AI Lab}
}

\author{Zihao Ye}
\email{yeziha@amazon.com}
\affiliation{%
  \institution{AWS Shanghai AI Lab}
}

\author{Jin Dong}
\email{jindon@amazon.com}
\affiliation{%
  \institution{AWS Shanghai AI Lab}
}

\author{Hao Xiong}
\email{xiongha@amazon.com}
\affiliation{%
  \institution{AWS Shanghai AI Lab}
}

\author{Zheng Zhang}
\email{zhaz@amazon.com}
\affiliation{%
  \institution{AWS Shanghai AI Lab}
}

\author{George Karypis}
\email{gkarypis@amazon.com}
\affiliation{%
  \institution{AWS AI}
}



\begin{abstract}
Knowledge graphs have emerged as a key abstraction for organizing information in diverse 
domains and their embeddings are increasingly used to harness their information in various 
information retrieval and machine learning tasks. However, the ever growing size of knowledge 
graphs requires computationally efficient algorithms capable of scaling to graphs with 
millions of nodes and billions of edges.
This paper presents \dglke, an open-source package to efficiently compute knowledge 
graph embeddings. \dglke introduces various novel optimizations that accelerate training 
on knowledge  graphs with millions of nodes and billions of edges using multi-processing, 
multi-GPU, and distributed parallelism. These optimizations are designed to increase data 
locality, reduce communication overhead, overlap computations with memory accesses, and 
achieve high operation efficiency.
Experiments on knowledge graphs consisting of over 86M nodes and 338M edges show that \dglke
can compute embeddings in 100 minutes on a EC2 instance with 8~GPUs and 30 minutes on
an EC2 cluster with 4~machines with 48~cores/machine. These results represent a 
$2\times \sim 5\times$ speedup over the  best competing approaches. \dglke is available
on https://github.com/awslabs/dgl-ke.
\end{abstract}


\maketitle

\section{Introduction}

Knowledge graphs (KGs) are data structures that store information about different 
entities (nodes) and their relations (edges). They are used to organize information 
in many domains such as music, movies, (e-)commerce, and sciences. A common approach 
of using KGs in various information retrieval and machine learning tasks is to compute 
knowledge graph embeddings (KGE)~\cite{wang2017knowledge,goyal2018graph}. 
These approaches embed a KG's entities and relation  types into a  $d$-dimensional space 
such that the embedding  vectors associated with  the  entities and the relation types 
associated with each edge satisfy a pre-determined  mathematical model. Numerous models 
for computing  knowledge graph embeddings have  been developed, such as 
TransE~\cite{transe},  TransR~\cite{transr} and  DistMult~\cite{distmult}. 

As the size of KGs has grown, so has the time required to compute their embeddings. 
As a result, a number of approaches and software packages have been developed that exploit 
concurrency in order to  accelerate the computations. Among them are GraphVite~\cite{graphvite}, 
which parallelizes the computations using multi-GPU training and Pytorch-BigGraph~(PBG)~\cite{pbg}, 
which uses distributed training to split the computations across a cluster of machines.
However, these approaches suffer from high data-transfer overheads and low computational 
efficiency. As a result, the time required to compute embeddings for large KGs is high.


In this paper we present various optimizations that accelerate KGE training on 
knowledge graphs with millions of nodes and billions of edges using 
multi-processing, multi-GPU, and distributed parallelism. These optimizations 
are designed to increase data locality, reduce communication overhead, overlap 
computations with memory accesses, and achieve high operation efficiency.

We introduce novel approaches of decomposing the computations across different 
computing units (cores, GPUs, machines) that enable massive parallelization while 
reducing write conflicts and communication overhead. 
The write conflicts are reduced by partitioning the processing associated with different 
relation types across the computing units as well as reducing data communication on 
multi-GPU training. 
The communication overhead is reduced by using  a min-cut-based graph partitioning 
algorithm (METIS~\cite{metis}) to  distribute the  knowledge graph across the machines. 
For entity embeddings, we introduce massive asynchronicity by having separate processes 
to compute the gradients of embeddings independently as well as allowing entity 
embedding updates overlapped with mini-batch computation. 
Finally, we use various negative sampling strategies to construct mini-batches with a small 
number of embeddings involved in a batch, which reduces data movement from memory to 
computing units (e.g., CPUs and GPUs).

We implement an open-source KGE package called \dglke that incorporates all of the optimization 
strategies to train KG embeddings on large KGs efficiently. The package is 
implemented with Python on top of Deep Graph Library (DGL)~\cite{dgl} along with
a C++-based distributed key-value store specifically designed for \dglke. We rely
on DGL 
to perform graph-related computation, such as sampling, and rely on existing deep 
learning frameworks, such as Pytorch~\cite{pytorch} and MXNet~\cite{mxnet}, to 
perform tensor computation. \dglke is available at https://github.com/awslabs/dgl-ke.

We experimentally evaluate the performance of \dglke on different knowledge graphs 
and compare its performance against GraphVite and Pytorch-BigGraph. Our experiments 
show that \dglke is able to compute embeddings whose quality is comparable to that of 
competing approaches at a fraction of their time. In particular, on knowledge graph 
containing over 86M nodes and 338M edges \dglke can compute the embeddings in 100 minutes 
on a EC2 instance with 8 GPUs and  30 minutes on an EC2 instance with 4 machines with 
48 cores/machine. These results represent  a $5\times$ and $2\times$ speedup over the 
time required by GraphVite and Pytorch-BigGraph, respectively.




\section{Background} \label{sec:background}
\paragraph{Definitions \& Notation}

A graph is composed of vertices and edges $G=(V, E)$, where $V$ is the set
of vertices and $E$ is the set of edges. A knowledge graph (KG) is a special type
of graph whose vertices and edges have types. It is a flexible data structure
that represents entities and their relations in a dataset. A vertex in a 
knowledge graph represents an entity and an edge represents a relation 
between two entities. The edges are usually in the form of triplets
$(h, r, t)$, each of which indicates that a pair of entities $h$ (\emph{head})
and $t$ (\emph{tail}) are coupled via a relation $r$.

Knowledge graph embeddings are low-dimensional representation of entities
and relations. These embeddings carry the information of the entities and
relations in the knowledge graph and are widely used in tasks, such as
knowledge graph completion and recommendation. Throughput the paper, we 
denote the embedding vector of head entity, tail entity and relation with $\mathbf{h}$, $\mathbf{t}$ and $\mathbf{r}$, respectively; all the embedding have the same dimension size of $d$.


\paragraph{Knowledge Graph Embedding (KGE) Models}

KGE models train entity embeddings and relation embeddings in a knowledge
graph. They define a score function on the triplets and optimize the function
to maximize the scores on triplets that exist in the knowledge graph and 
minimize the scores on triplets that do not exist.

Many score functions have been defined to train knowledge graph embeddings
\cite{kge} and Table~\ref{tbl:kge} lists the ones used by the KGE models
supported by \dglke. TransE and TransR are two representative translational 
distance models, where we use L1 or L2 to define the distance. DistMult, 
ComplEx, and RESCAL are semantic matching models that exploit similarity-based 
scoring functions. Some of the models are much more computationally expensive 
than other models. For example, TransR is $d$ times more computationally
expensive than TransE because TransR has additional matrix multiplications
on both head and tail entity embeddings, instead of just element-wise operations
on embeddings in TransE.

\begin{table}
\begin{center}
\footnotesize
\caption{Knowledge graph models.
$M_r$ is a relation-specific projection matrix. TransE uses L1 or L2
norm in its score function.}
\label{tbl:kge}
\begin{tabular}{lc}
\toprule
Models & score function $f(\mathbf{h}, \mathbf{r}, \mathbf{t})$ \\ 
\midrule
TransE \cite{transe} & $-||\mathbf{h}+\mathbf{r}-\mathbf{t}||_{1/2}$ \\
TransR \cite{transr} & $-||M_r \mathbf{h} + \mathbf{r} - \mathbf{M_r} \mathbf{t}||_2^2$ \\
DistMult \cite{distmult} & $\mathbf{h}^\top \mbox{diag}(\mathbf{r}) \mathbf{t}$ \\
ComplEx \cite{complex} & $\mbox{Real}(\mathbf{h}^\top \diag(\mathbf{r}) \mathbf{\bar{t}})$ \\
RESCAL \cite{rescal} & $\mathbf{h}^\top \mathbf{M_r} \mathbf{t}$ \\
RotatE \cite{rotate} & $-||\mathbf{h}\circ\mathbf{r}-\mathbf{t}||^2$\\ 
\bottomrule
\end{tabular}
\normalsize
\end{center}
\end{table}

To train a KGE model, we define a loss functions on a set of positive and
negative samples from the knowledge graph. Two loss functions are commonly 
used. The first is the a logistic loss given by 

$$\mbox{minimize} \sum_{\mathbf{h},\mathbf{r},\mathbf{t} \in \mathbb{D}^+ \cup \mathbb{D}^-} \log(1+\exp(-y \times f(\mathbf{h}, \mathbf{r}, \mathbf{t}))),$$

\noindent where $\mathbb{D}^+$ and $\mathbb{D}^-$ are the positive and negative sets of 
triplets, respectively and $y$ is is the label of a triplet, $+1$ for positive and $-1$ 
for negative. The second is the  pairwise ranking loss given by

$$\mbox{minimize} \sum_{\mathbf{h},\mathbf{r},\mathbf{t} \in \mathbb{D}^+} \sum_{\mathbf{h}',\mathbf{r}',\mathbf{t}' \in \mathbb{D}^-} \max(0, \gamma - f(\mathbf{h}, \mathbf{r}, \mathbf{t}) + f(\mathbf{h}',\mathbf{r}',\mathbf{t}')).$$

A common strategy of generating negative samples is to corrupt a triplet by replacing its 
head entity or tail entity with entities sampled from the graph with some heuristics to 
form a negative sample $(h, r, t')$ or $(h', r, t)$, where $h'$ and $t'$ denote the 
randomly sampled entities. Potentially, we can corrupt the relation in a triplet. In 
this work, we only corrupt entities to generate negative samples.

\paragraph{Mini-batch training and Asynchronous updates} 

A KGE model is typically trained in a mini-batch fashion. We first sample a mini-batch
of $b$ triplets $(h, r, t)$ that exist in the knowledge graph. The mini-batch training
is sparse because a batch only involves in a small number of entity embeddings and 
relation embeddings. We can take advantage of the sparsity and train KGE models 
asynchronously with sparse gradient updates~\cite{hogwild}. That is, we sample multiple 
mini-batches independently, perform asynchronous stochastic gradient descent (SGD) on 
these mini-batches in parallel and only update the embeddings involved in the mini-batches. 
This training strategy maximizes parallelization in mini-batch training but may lead to conflicts in updating gradients. When two mini-batches run 
simultaneously, they may use the same entity or relation embeddings. In this case, the 
gradient of the embeddings is computed based on the stale information, which results 
in a slower convergence or not converging to the same local minimum.

\section{Methods}

A naive implementation of KGE training results in low computation-to-memory density
for many KGE models, which prevents us from using computation resources efficiently.
When performing computation on a batch, we need to move a set of entity and relation embeddings
to computation resources (e.g., CPUs and GPUs) from local CPU memory or remote machines.
For example, for a mini-batch with $b$ positive triplets, $k$ negative triplets, and 
$d$-dimensional  embeddings, both the computational and data movement complexity of 
TransE is $O(b  d  (k + 1))$, resulting in a computational density of $O(1)$. Given that 
computations are faster than memory accesses, reducing data movement is key to 
achieving efficient KGE training.

In addition, we need to take advantage of parallel computing resources. 
This includes multi-core CPUs,
GPUs and a cluster of machines. Our training algorithm needs to allow massive 
parallelization while still minimizing conflicts when updating embeddings in parallel.

In this work, we implement \dglke on top of DGL~\cite{dgl}, 
completely with Python. It relies on DGL for graph computation, such as sampling,
and relies on deep learning frameworks, such as Pytorch and MXNet, for tensor operations.

\subsection{Overview}\label{sec:overview}

\begin{figure*}
\centering
\includegraphics[scale=0.50]{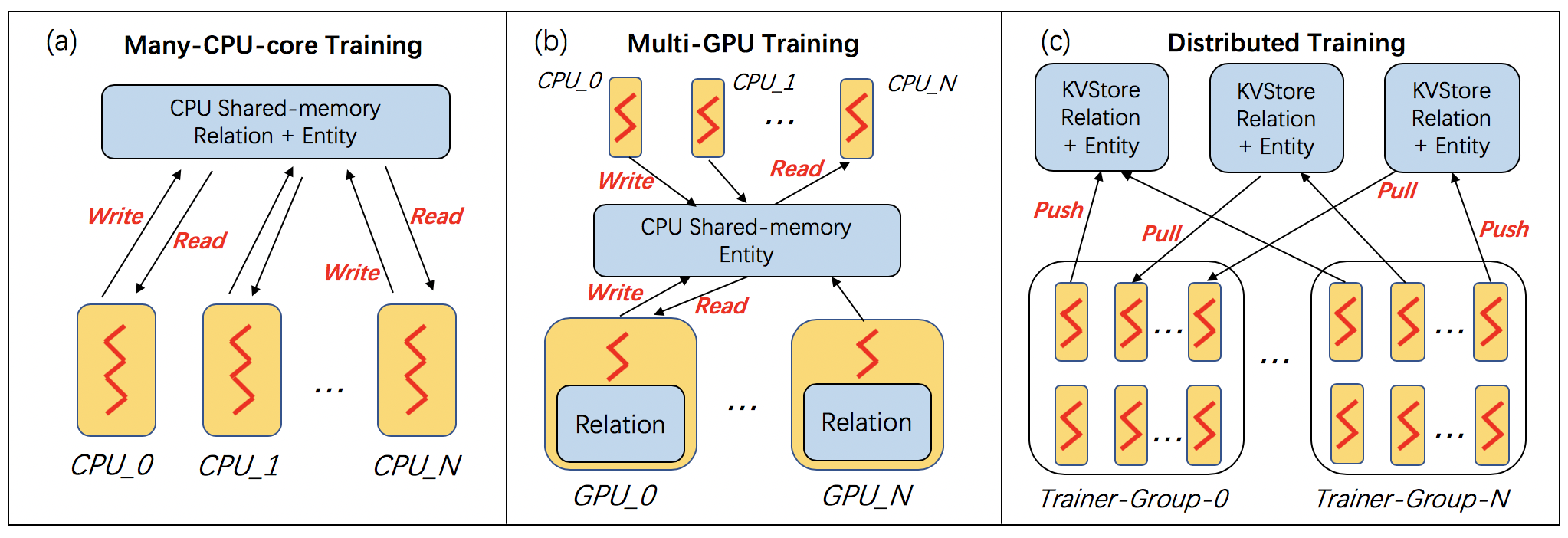}
\caption{The optimized data placement of \dglke in three different parallel hardware.}
\label{fig:arch}
\end{figure*}

\dglke provides a unified implementation for efficient KGE training on different hardware.
It optimizes for three types of hardware configurations: (i) many-core CPU machines, 
(ii) multi-GPU machines, and (iii) a cluster of CPU/GPU machines. In each type of the 
hardware, \dglke parallelizes the training with multiprocessing to fully utilize the 
parallel computation power of the hardware.

For all different hardware configurations, the training process starts with a 
preprocessing step to partition a knowledge graph and follows with mini-batch 
training. The partitioning step assigns a disjoint set of triplets in a knowledge 
graph to a process so that the process performs mini-batch training independently.

The specific steps performed during each mini-batch are:

\begin{enumerate}
    \item Samples triplets from the local partition that belongs to a process to 
    form a mini-batch and constructs negative samples in the mini-batch.
    
    \item Fetches entity and relation embeddings that are involved in the mini-batch 
    from the global entity and relation embedding tensors.
    
    \item Performs forward computation and back-propagation on the embeddings fetched
    in the previous step in order to compute the gradients of the embeddings.
    
    \item Applies the gradients to update the embeddings involved in the mini-batch.
    This step requires to apply an optimization algorithm to adjust the gradients and 
    write the gradients back to the global entity and relation embedding tensors.
\end{enumerate}

KGE training on a knowledge graph involves two types of data: the knowledge graph structure
and the entity and relation embeddings. As illustrated in Figure~\ref{fig:arch}, we deploy different data placement 
for different hardware configurations. 
In many-core CPU  machines, \dglke keeps the knowledge graph structure as well as 
entity and relation  embeddings in shared CPU memory accessible to all processes. 
A trainer process reads  the entity and relation embeddings from the global 
embeddings directly through shared memory.
In multi-GPU machines, \dglke keeps the knowledge graph structure and entity embeddings
in shared CPU memory because entity embeddings are too large to fit in GPU memory.
It may place relation embeddings in GPU memory to reduce data communication.
As such, a trainer process reads entity embeddings from CPU shared memory
and reads relation embeddings directly from GPU memory.
In a cluster of machines, \dglke implements a C++-based distributed key-value store (KVStore) to 
store both entities and relation embeddings. The KVStore partitions the entity
embeddings and relation embeddings automatically and strides them across all 
KVStore servers. A trainer process accesses embeddings from distributed KVStore with 
the pull and push API. We partition the knowledge graph structure and each trainer 
machine stores a partition of the graph. The graph structure of the partition 
is shared among all trainer processes in the machine.

The rest of this section describes various optimization techniques that 
we developed in \dglke:
graph partitioning in the preprocessing step (Section~\ref{sec:partition}),
negative sampling (Section~\ref{sec:neg_sample}),
data access to
relation embeddings (Section~\ref{sec:rel_part}), and finally
applying gradients to the global embeddings (Section~\ref{sec:overlap}).




\subsection{Graph partitioning} \label{sec:partition}
In distributed training, we partition the graph structure and embeddings and store
them across the machines of the cluster. During the training, each machine may need to read
entity and relation embeddings from other machines to construct mini-batches.
The key of optimizing distributed training is to reduce communication
required to retrieve and update entity and relation embeddings.


\begin{figure}
    \centering
	\includegraphics[width=6cm]{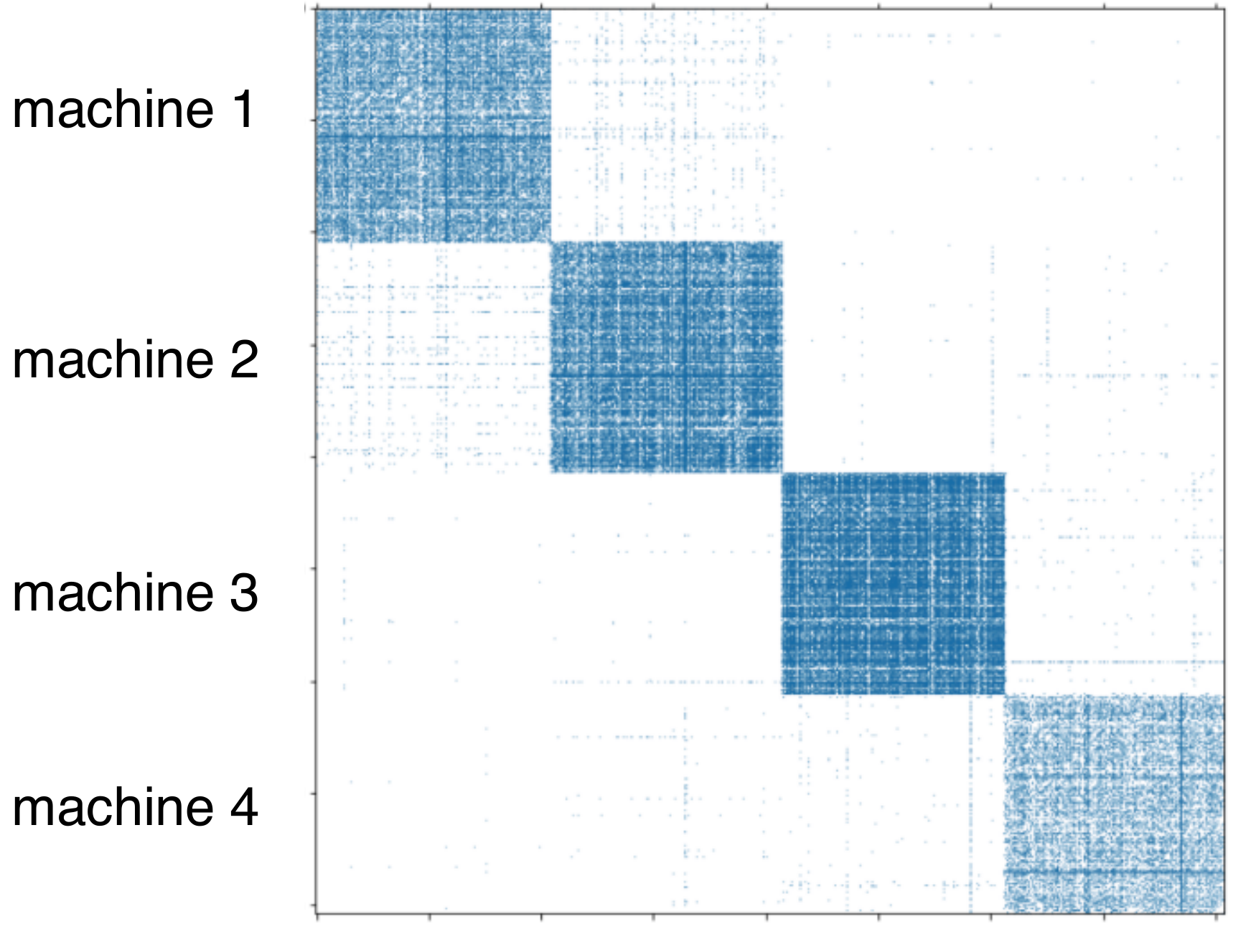}
    \caption{Adjacent matrix of a large graph after applying METIS partitioning, indexed by machine partition. Note that majority of the edges fall within a partition. As a result,
    the adjacency matrix has majority of non-zeros lying on the diagonal blocks.}
    \label{fig:metis}
\end{figure}

To reduce the communication caused by entity embeddings in a batch,
we deploy METIS partitioning \cite{metis} on the knowledge graph in the preprocessing step.
For a cluster of $P$ machines, we split the graph into $P$ partitions so that
we assign a METIS partition (all entities and triplets incident to the entities) to a machine
as shown in Figure~\ref{fig:metis}. With METIS partitioning, the majority of
the triplets are in the diagonal blocks. We co-locate the embeddings of
the entities with the triplets in the diagonal block by specifying a proper
data partitioning in the distributed KVStore.
When a trainer process samples triplets in the local partition,
most of the entity embeddings accessed by the batch fall in the local partition
and, thus, there is little network communication to access entity embeddings
from other machines.

\subsection{Negative sampling} \label{sec:neg_sample}

KGE training samples triplets to form a batch and construct a large number of
negative samples for each triplet in the batch.
For all different hardware, \dglke performs sampling on CPUs and offloads
the entire sampling computation to DGL for efficiency.
If we construct negative samples independently for each triplet, a mini-batch will contain
many entity embeddings, which results in accessing many embeddings.

We deploy a joint negative sampling to reduce the  
number of entities involved in a mini-batch. In this approach, instead of  independently 
corrupting every triplet $k$ times, we group the triplets into sets of size 
$g$ and corrupt them together. For example, when corrupting the tail entities of a set,  we 
uniformly sample $k$ entities to replace the tail entities of that set. We corrupt the head 
entities in a similar fashion. This negative sampling strategy introduces two benefits. 
First, it reduces the number of entities involved in a mini-batch, resulting in 
a smaller amount of data access. For a $d$-dimensional embedding, each mini-batch 
of size $b$ now only needs to access $O(bd + bkd/g)$ instead of $O(b d  (k + 1))$ words of 
memory. When $g$ grows as large as $b$, the amount of data accessed by this negative sampling 
is about $b$ times smaller ($b$ is usually in the order of 1000). This benefit is more 
significant in multi-GPU training because we store entity embeddings in CPU memory and send the entity  embeddings to the GPUs in every mini-batch.
Second, it allows us to replace the original computation with more efficient tensor operations.
Inside a group of negative samples, head entities and tail entities are densely connected. 
We now divide the computation of a score function on a negative sample into two parts. 
For example, the score function of TransE\_l2, $-||\mathbf{h}+\mathbf{r}-\mathbf{t}||_2$, is 
divided into $\mathbf{o}=\mathbf{h}+\mathbf{r}$  and $-||\mathbf{o}-\mathbf{t}'||_2$. 
The vector $\mathbf{o}$ is computed as before because there are only $b$ pairs of 
$\mathbf{h}$ and $\mathbf{r}$. The computation of $-||\mathbf{o}-\mathbf{t}'||_2$ is converted  into a generalized matrix multiplication, which can be performed using highly optimized math libraries. There are $b k$ pairs of $\mathbf{o}$ and $\mathbf{t}'$. 

We also deploy non-uniform negative sampling with a probability  proportional to the 
degree of each entity (PBG uses a similar strategy). On a large knowledge graph, uniform negative sampling results 
in easy negative samples~\cite{kotnis2017analysis}. One way of constructing \emph{harder} 
negative  samples is  to corrupt a triplet with entities sampled proportional to the entity 
degree. In order to do this efficiently, instead of 
sampling entities from the entire graph, we construct negative samples with the entities 
that are already in the mini-batch. This is done by uniformly sampling some of the 
mini-batch's triplets and connecting the sampled head (tail) entities with the tail (head) 
entities  of the mini-batch's triplets to construct the negative samples. Note that this 
uniform triplet sampling approach leads to an entity sampling approach that is proportional 
to the entity degree in the mini-batch. In practice, we combine these negative  samples 
with uniformly negative samples to form the full set of negative samples for a mini-batch.


In the distributed training, we sample entities from the local METIS partition
to corrupt triplets in a mini-batch to minimize the communication caused by
negative samples. This ensures that
negative samples do not increase network communication. This strategy in general results in harder
negative samples. The corrupted head/tail entities sampled from the local METIS partition
are topologically closer to the tail/head entities of the triplets in the batch.

\subsection{Relation partitioning} 
\label{sec:rel_part}




Both GraphVite and PBG treat relation embeddings as dense model weights.
As a result, for each mini-batch they incur the cost of retrieving them and updating them.
If the number of relations in the knowledge graph is small, this is close to optimal and 
does not impact the performance. However, when the knowledge graph has a large number 
of relations (greater than the mini-batch size; $\approx 1000$), the number 
of distinct relations in each mini-batch will be a subset of them and as such, treating them
as dense model weights will result in unnecessary data access/transfer overheads.
To address this limitation, \dglke performs sparse relation embedding reads and sparse 
gradient updates on relation embeddings. This significantly reduces the amount of data 
transferred in multi-processing, multi-GPU, and distributed training.

To further reduce the amount of access to relation embeddings
in a mini-batch, \dglke decomposes the computations among the computing 
units by  introducing a novel relation partitioning approach. This relation partitioning 
tries  (i) to equally distribute the triplets  and the relations among the partitions and 
(ii) to minimize the number of distinct relations that are assigned to each partition 
as a result of (i). The first goal  ensures that the computational and memory requirements 
are balanced across the computing units, whereas the second goal ensures that the 
relation-related data that needs to be  transferred is minimized. 
In order to derive such a relation partitioning, we use the following fast greedy 
algorithm. We sort the relations based on their frequency in non-increasing order. 
We iterate over the sorted relations and greedily assign a relation to the partition 
with the smallest number of triplets so far. This strategy usually results 
in balanced partitioning while ensuring that each relation belongs to only one partition.
However, the above algorithm will fail to produce a balance partitioning when the 
knowledge graph contains relations that are very frequent. In such cases, the number
of triplets for those relations may exceed the partition size. To avoid load imbalance,
we equally split the most common relations across all partitions.
After relation partitioning, we assign a relation partition to a computing unit.
This ensures that the majority of relation embeddings are updated by only one process at 
a time. This optimization applies to many-CPU-core training and multi-GPU training.

A potential drawback of relation partitioning is that it restricts the relations
that may appear inside a mini-batch. This reduces the randomization of stochastic 
gradient descent, which can impact the quality of the embeddings. To tackle this 
problem, we introduce randomization in the partitioning algorithm and at the start 
of each epoch we compute a somewhat different relation partitioning.

When we use relation partitioning in  multi-GPU training, we store all relation embeddings 
on GPUs and update relation embeddings in GPUs locally. This is particularly important 
for KGE models  with large model weights on relations, such as TransR and RESCAL. 
Take TransR for  an example. It has an entity projection matrix on each relation, 
which is much  larger than a relation embedding. Moving them to CPU is the bottleneck 
of the  entire computation. If we keep all of these projection matrices in GPUs, the 
communication overhead drops from $O(bd^2)$ to $O(bd)$, which is significantly 
smaller than the naive solution, usually in the order of $100$ times smaller.



\subsection{Overlap gradient update with batch processing} \label{sec:overlap}

In multi-GPU training, some of the steps in a mini-batch computation run on
CPUs while the others run on GPUs. When we run them in serial in a process,
the GPU remains idle when the CPU writes the gradients. To avoid GPU idling,
we overlap entity embedding update with the batch computation in the next mini-batch.
This allows us to overlap the computation in CPUs and GPUs. Note that even though 
this approach can potentially increase the staleness of  the embeddings used in 
a mini-batch, the likelihood of that happening is small for knowledge graphs 
with a sufficiently large number of entities relative to the number of training 
processes.

To perform this optimization, we split the gradient updates into two parts: one 
involving relation embeddings, which are updated by the trainer process, and the 
other involving the entity embeddings, which are off-loaded to a dedicated 
gradient update process for each trainer process. Once the trainer process 
finishes writing the relation gradients, it proceeds to the next mini-batch, 
without having to wait for the writing of the entity gradients to finish.
Our experiments show that overlapping gradient updates provide 40\% speedup 
for most of the KGE models on Freebase.

\subsection{Other optimizations} \label{sec:other_opts}

\paragraph{Periodic synchronization among processes}
When training KGE models with multiprocessing completely independently, different
processes may run at a different rate, which results in inconsistent model accuracy.
We observe that the trained embeddings sometimes have much worse accuracy at some runs.
As such, we add a synchronization barrier among all training processes after a certain
number of batches to ensure that all processes train roughly at the same rate.
Our observation is that the model can be trained stably if processes synchronize
after every few thousand batches.

\paragraph{Distributed Key-Value store} 

In \dglke, we implement a KVStore for model synchronization with efficient C++ 
back-end. It uses three optimizations that are designed specifically for distributed 
KGE training. 
First, because the relations in some knowledge graphs have a long-tail distribution, 
it reshuffles the relation embeddings in order to avoid single hot-point of KVStore.
Second, \dglke uses local shared-memory access instead of network communication 
if the worker processes and KVStore processes are on the same machine. This 
optimization can significantly reduce networking overhead especially on METIS 
graph partition.
Third, it launches multiple KVStore servers in a single machine to
parallelize the computation in KVStore.
All KVStore servers inside a machine share embeddings via local shared-memory. Finally, similar to the optimization we used in multi-GPU training, the gradient communication and local gradient computation will be overlapped in KVStore.


\section{Related Work}
There are a few packages that have been developed to compute
embeddings of knowledge graphs efficiently and scale to large
knowledge graphs. 

OpenKE~\cite{openke} is one of the first packages for training knowledge 
graph embeddings and provides a large list of models. However, it is 
implemented entirely in Python and cannot scale to very large graphs.

Pytorch-BigGraph~(PBG)~\cite{pbg} is developed with an emphasis on scalability to 
large graphs and distributed training on a cluster of machines. The package does 
not support GPU training. Although PBG and \dglke share similar negative sampling
strategies, PBG applies different strategies for distributed training. It randomly 
divides  the adjacency matrix of the graph into 2D blocks and assigns blocks to 
each machine based on a schedule that avoids conflicts with respect to the entity 
embeddings. It treats entity embeddings as sparse model weights and relation
embeddings as dense model weights. The random 2D partitioning along with the use of 
dense model weights for relation embeddings results in a large amount of communication, 
especially for knowledge graphs with many relations.

GraphVite \cite{graphvite} focuses on multi-GPU training
and does not support distributed training. When it trains a large knowledge 
graph, it keeps embeddings on CPU memory. It constructs a subgraph, moves 
all data in the subgraph to the GPU memory and performs many mini-batch 
training steps on the subgraph. This method reduces data movement between 
CPUs and GPUs at the cost of increasing the staleness of the embeddings, 
which usually results in slower convergence.

\section{Experimental Methodology}
\label{sec:exp}

\dglke is implemented in Python and relies on PyTorch for tensor operations, as is the case in PBG, whereas GraphVite is done mostly in C++ with a Python wrapper. We report \dglke performance in two broad section: 
(i) on multi-GPU in  section \ref{eval-multi-gpu}, many-core CPU in section 
\ref{eval-many-core} and distributed training in Section~\ref{eval-dist}, 
(ii) against  GraphVite~\cite{graphvite} and PBG~\cite{pbg} in Section~\ref{eval-compare} 
on identical hardware. 


\subsection{Hardware platform} 

We conduct our evaluation on EC2 CPU and GPU instances, including GraphVite and PBG; see Table~\ref{table:eval-hard} for machine configurations.

\begin{table}
\begin{center}
\footnotesize
\caption{Evaluation hardware configuration.}
\label{table:eval-hard}
\begin{tabular}{lll}
\toprule
EC2 Type & Hardware Config & Eval Section \\
\midrule
r5dn.24xlarge & 2x24 cores, 700GB RAM, 100Gbps network & sec 6.2, 6.3 \\
p3.16xlarge & 2x16 cores, 500GB RAM, 8 V100 GPUs & sec 6.1 \\
\bottomrule
\end{tabular}
\normalsize
\end{center}
\end{table}

\begin{table}
\begin{center}
\footnotesize
\caption{Knowledge graph datasets.}
\label{datasets}
\begin{tabular}{lrrrl}
\toprule
Dataset & \# Vertices & \# Edges & \#Relations &  Systems\\
\midrule
FB15k~\cite{transe}      & 14,951     & 592,213     & 1345   & \dglke, GraphVite\\
WN18~\cite{transe}       & 40,943     & 151,442     & 18     & \dglke, GraphVite\\
Freebase~\cite{freebase} & 86,054,151 & 338,586,276 & 14,824 & \dglke, PBG\\
\bottomrule
\end{tabular}
\normalsize
\end{center}
\end{table}

\subsection{Datasets} 

We used three datasets to evaluate and compare the performance of \dglke against that of GraphVite and PBG. Table~\ref{datasets} shows various statistics for these datasets. FB15k and Freebase were derived from the Freebased Knowledge Graph~\cite{freebase}, whereas WN18 was derived from WordNet~\cite{wordnet}. The FB15k and WN18 datasets are standard benchmarks for evaluating KGE methods. The Freebase dataset corresponds to complete Freebase Knowledge Graph. All datasets are downloaded from \cite{kg_data}.

\subsection{Evaluation methodology} 

We evaluated the performance of the different KGE models and methods using a link (relation)-prediction task. In order to train and evaluate the models, we split each dataset into training, validation, and test subsets. For FB15k and WN18, we used the same splits that were used in previous evaluations~\cite{rotate} (available in~\cite{kg_data}). Freebase is split with 5\% of the triplets for validation, 5\% for test, and the remaining 90\% for training (also available in~\cite{kg_data}).

We performed the link-prediction task using two different protocols. The first, which was used for FB15k and WN18, works as follows. For each triplet $(h, r, t)$ in the validation/test set, referred to as \emph{positive triplet}, we generated all possible triplets of the form $(h', r, t)$ and $(h, r, t')$ by corrupting the head and tail entities. We then removed from them any triplets that already exist in the dataset. The set of triplets that remained form the \emph{negative triplets} associated with the initial positive triplet. We then used the score function of the model in question (Table~\ref{tbl:kge}) to score the triplets.
The second protocol, which was used for Freebase, is similar to the first one with the following two differences: (i) we use only 2000 negative triplets; 1000 sampled uniformly from the entire set of negative samples and 1000 sampled proportionally to the degree of the corrupted entities; and (ii) we did not remove from the 2000 negative triplets any triplets that are in the dataset. Note that the reason for the second protocol was due to the size of Freebase, which made the first protocol computationally expensive.

We assessed the performance by using the standard metrics~\cite{pbg} of Hit@$k$ (for $k\in \{1, 3, 10\}$), Mean Rank (MR), and Mean Reciprocal Rank (MRR). All these metrics are derived by comparing how the score of the positive triplet relates to the scores of its associated negative instances. For a positive triplet $i$, let $S_i$ be the list of triplets containing $i$ and its associated negative triplets ordered in a non-increasing score order, and let $rank_i$ be $i$th position in $S_i$. Given that,  Hit@$k$ is the average number of times the positive triplet is among the $k$ highest ranked triplets; MR is the average rank of the positive instances, whereas MRR is the average reciprocal rank of the positive instances.
Mathematically, they are defined as
$$\mbox{Hit@}k=\frac{1}{Q}\sum_{i=1}^Q \mathbbm{1}_{rank_i \le k},$$
$$\mbox{MR}=\frac{1}{Q} \sum_{i=1}^Q rank_i,$$
and
$$\mbox{MRR}=\frac{1}{Q} \sum_{i=1}^Q \frac{1}{rank_i},$$
where $Q$ is the total number of positive triplets and $\mathbbm{1}_{rank_i \le k}$ is 1 if $rank_i \le k$, otherwise it is 0. Note that Hit@$k$ and MRR are between 0 and 1, whereas MR ranges from 1 to the $\sum_i^Q |S_i|$.

\subsection{Software environment}

We run Ubuntu 18.04 on all EC2 instances, where the Python version is 3.6.8 and Pytorch version is 1.3.1. On GPU instances, the CUDA version is 10.0. When comparing the performance of \dglke against that of GraphVite and PBG, we use GraphVite v0.2.1 downloaded from Github on November 12 2019 and PBG downloaded from their Github repository on October 15 2019. All frameworks use the same Pytorch version.

\subsection{Hyperparameters} 

For the FB15k and WN18 and all methods (\dglke and GraphVite) we performed an extensive hyper-parameter search and report the results that achieve the best performance in terms of MRR, as we believe it is a good measure to assess the overall performance of the methods. 
%
Due to the size of Freebase, we only report results for a single set of hyper-parameter values. We use the hyperparameters that perform the best on FB15k for Freebase.

To ensure that the accuracy results are comparable, all methods used exactly the same test set and evaluation protocols described in the previous section.

%

\section{Results}
\label{sec:results}

\subsection{Multi-GPU training} \label{eval-multi-gpu}

Both memory and computing capacity on a multi-GPU machine have a diverse set 
of characteristics, which make the various optimizations described in 
Sections~\ref{sec:neg_sample}--\ref{sec:other_opts} relevant. A detailed 
evaluation of these optimizations follows.


\subsubsection{Negative sampling}

\begin{figure}%
    \centering
    \includegraphics[width=8.5cm]{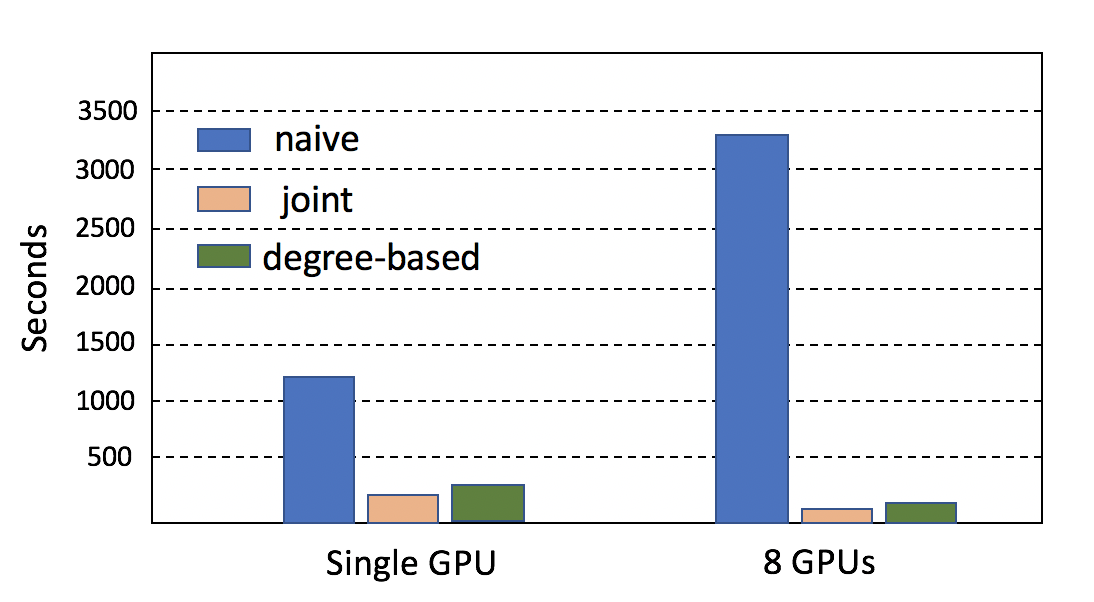}
    \caption{The effect of negative sampling in GPU training on FB15k.}%
    \label{fig:multigpu_neg_sample}%
\end{figure}

Joint negative sampling shown in Section~\ref{sec:neg_sample} has
two effects: (i) enable more efficient tensor operators and (ii) reduce
data movement in multi-GPU training. Figure~\ref{fig:multigpu_neg_sample} 
shows the result. To illustrate the speedup of using more efficient tensor 
operators, we run the TransE model on FB15k with all data in a single GPU.
Joint negative sampling gives about $4 \times$ speedup.
To illustrate the speedup of reducing data movement,
we run the TransE model on FB15k in 8 GPUs, where the entity
embeddings are stored in CPU memory. Join negative sampling
gets much larger speedup, e.g., about $40 \times$, because naive sampling requires
swapping many more entity embeddings between CPU and GPU than joint 
negative sampling and data communication becomes the bottleneck.

\subsubsection{Degree-based negatvie sampling}

Although degree-based negative sampling does not speed up training,
it improves the model accuracy (Table ~\ref{table:degree_sample})
on Freebase. This suggests that non-uniform negative sampling to generate
``hard'' negative samples is effective, especially on large knowledge graphs.

\begin{table}
\begin{center}
\footnotesize
\caption{The performance of KGE models on Freebase with and without
degree-based negative sampling with eight GPUs.}
\label{table:degree_sample}
\begin{tabular}{lcccccc}
\toprule
      & \multicolumn{2}{c}{TransE} & \multicolumn{2}{c}{ComplEx} & \multicolumn{2}{c}{DistMult} \\
  & with & w/o & with & w/o & with & w/o \\
\midrule
Hit@10 & 0.834 & 0.783 & 0.777 & 0.638 & 0.742 & 0.731 \\ 
Hit@3  & 0.773 & 0.675 & 0.741 & 0.564 & 0.698 & 0.697 \\ 
Hit@1  & 0.689 & 0.527 & 0.677 & 0.485 & 0.639 & 0.652 \\ 
MR     & 41.16 & 43.99 & 108.43 & 162.74 & 123.10 & 128.91 \\ 
MRR    & 0.743 & 0.619 & 0.716 & 0.539 & 0.678 & 0.682 \\ 
\bottomrule
\end{tabular}
\normalsize
\end{center}
\end{table}

\begin{figure}
\centering
\includegraphics[scale=0.3]{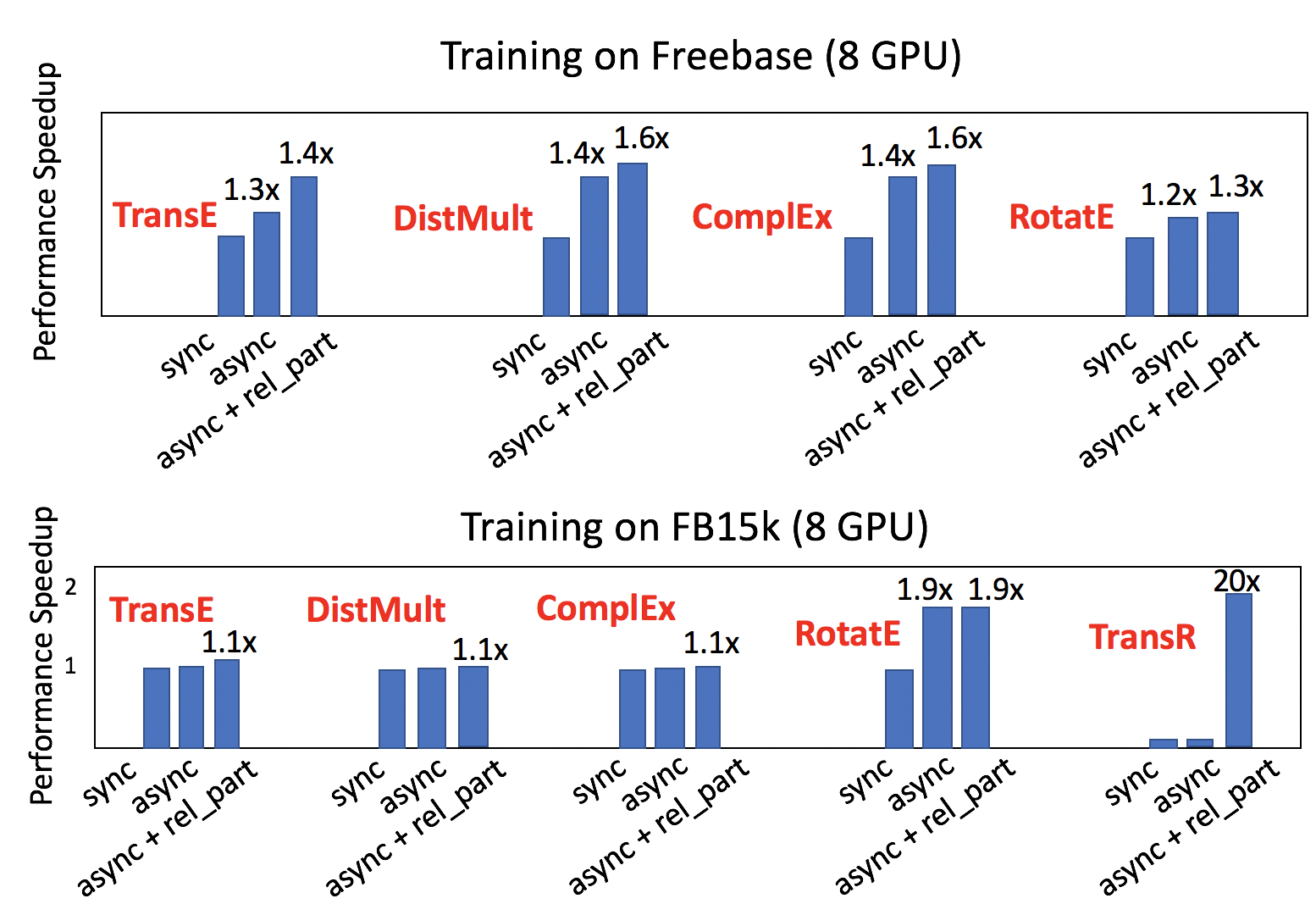}
\caption{Speedup of different optimizations on multi-GPU.}
\label{fig:opt-speedup}
\end{figure}

\subsubsection{Overlap gradient update with batch computation}

This technique overlaps the computation of GPUs and CPUs to speed up
the training. Figure \ref{fig:opt-speedup} shows the speedup of using 
this technique (comparing \textbf{sync} and \textbf{async}) on 
FB15k and Freebase. It has limited speedup on small knowledge graphs
for some models, but it has roughly 40\% speedup on Freebase for almost
all models. The effectiveness of this optimization depends on
the computation time in CPUs and GPUs. Large knowledge graphs, such as 
Freebase, requires hundreds of GBytes to store the entire entity embeddings
and suffers from slow random memory access during entity embedding 
update. In this case, overlapping the CPU/GPU computation benefits a lot.

\subsubsection{Relation partitioning}

After relation partitioning, we pin relation embeddings (and projection
matrices) in each partition inside certain GPU, which reduces the data movements 
between CPUs and GPUs. The speedup is highly related to the model size and 
the number of relations in the dataset.
Figure \ref{fig:opt-speedup} shows the speedup of using relation partitioning in 
multi-GPU training (comparing \textbf{async} and \textbf{async + rel\_part} bar) 
on FB15k and Freebase.
For example, relation partitioning has significant speedup on TransR because
the relation-specific projection matrices result in a large amount of data
communication between CPU and GPU. Even for models with only relation
embeddings, relation partitioning in general gets over 10\% speedup.



\subsubsection{Overall speed and accuracy}

\begin{figure}%
    \centering
    \subfloat[Training on FB15k]{{\includegraphics[width=8cm]{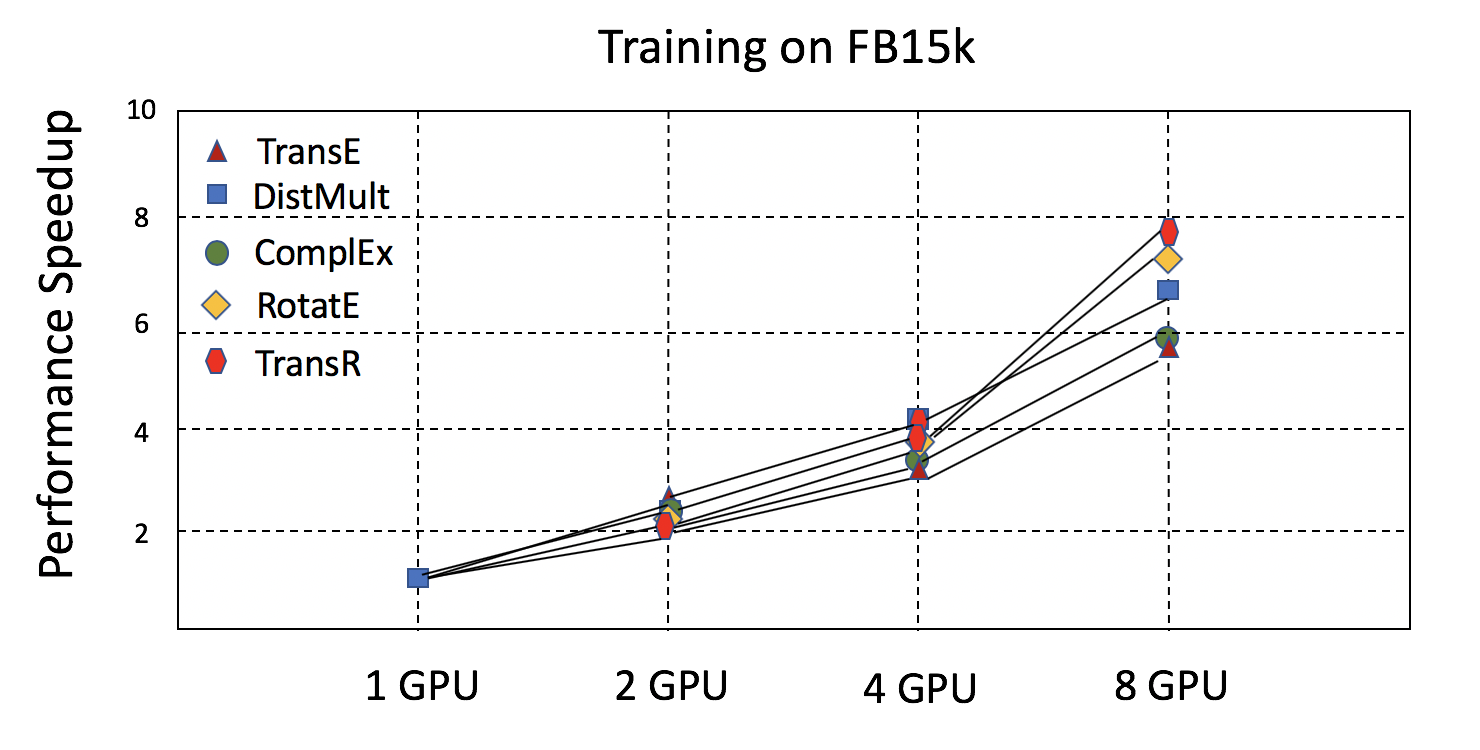} }}%
    \qquad
    \subfloat[Training on Freebase]{{\includegraphics[width=8cm]{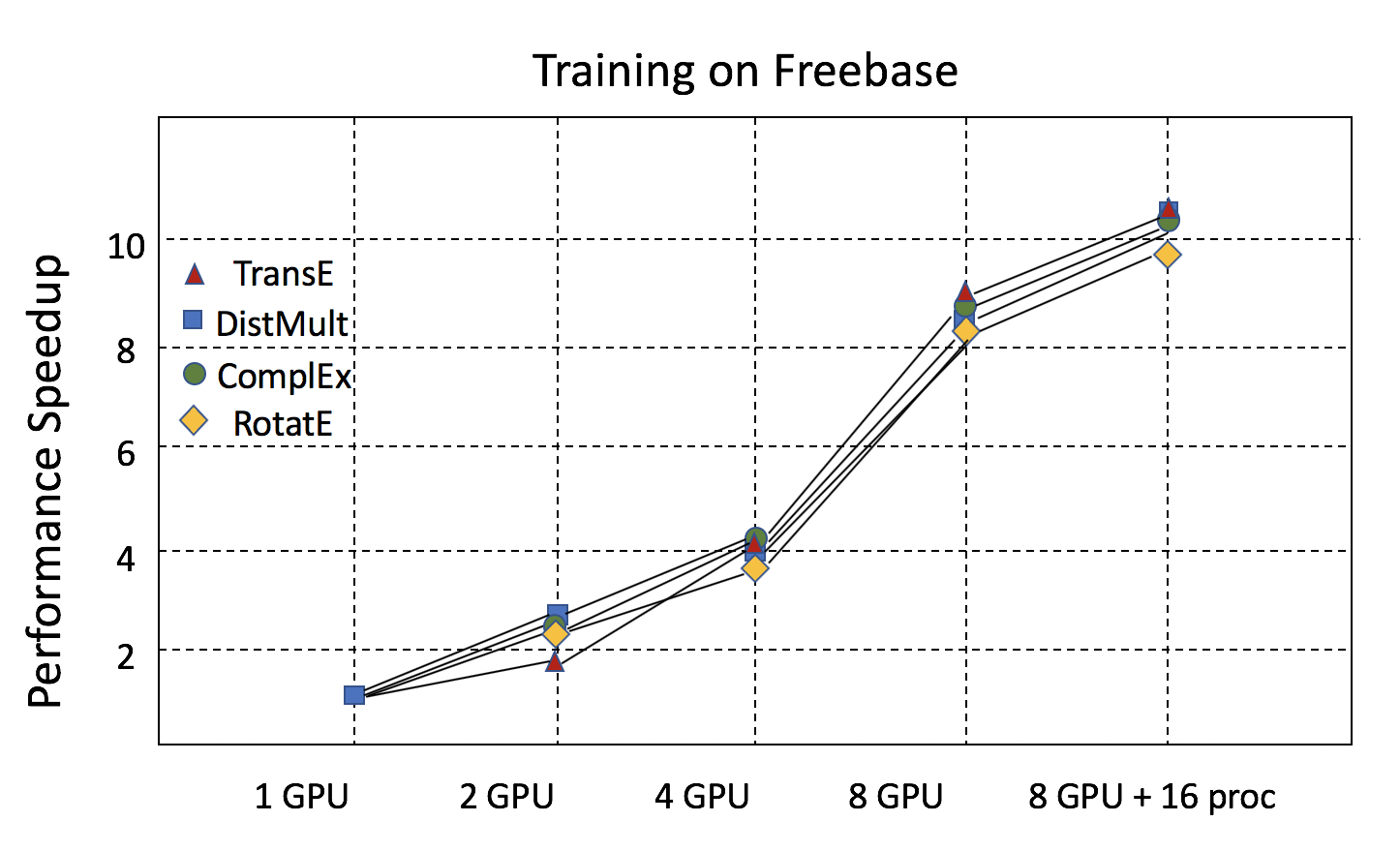} }}%
    \caption{Speedup of multi-GPU training.}%
    \label{fig:speedup}%
\end{figure}

After deploying all of the optimizations evaluated above, we measure
the speedup of \dglke with multiple GPUs on both FB15k and Freebase.
Figure \ref{fig:speedup} shows
that \dglke accelerates training almost linearly with multiple GPUs.
On Freebase, \dglke further speeds up by running 16 processes on 8 GPUs.
By running two processes on each GPU, we better utilize
the computation in GPUs and PCIe buses by overlapping computation and 
data communication between CPUs and GPUs.

With all these techniques, we train KGE models efficiently.
For small knowledge graphs, such as FB15k, \dglke trains most of KGE
models, even as complex as RotatE and TransR, within a few minutes.
For large knowledge graphs, such as Freebase, \dglke trains many of KGE
models around one or two hours and trains more complex models within a 
reasonable time, for example we train TransR in about 8 hours using 8 GPUs.


\begin{table*}[t]
\begin{center}
\setlength{\tabcolsep}{5pt}
\footnotesize
\caption{The overall performance of \dglke after various 
optimizations with 8 GPUs on FB15k.}
\label{table:overall_opt_acc_fb15k}
\begin{tabular}{lcccccccccc}
\toprule
       & \multicolumn{2}{c}{TransE\_l1} & \multicolumn{2}{c}{DistMult} & \multicolumn{2}{c}{ComplEx} &
\multicolumn{2}{c}{RotatE} &
\multicolumn{2}{c}{TransR} \\
  & 1GPU & Fastest & 1GPU & Fastest & 1GPU & Fastest & 1GPU & Fastest & 1GPU & Fastest \\
\midrule
Hit@10 & 0.860 & 0.857 & 0.884 & 0.879 & 0.892 & 0.884 & 0.885 & 0.874 & 0.820 & 0.815 \\
Hit@3 & 0.775 & 0.765 & 0.806 & 0.796 & 0.838 & 0.823 & 0.819 & 0.804 & 0.742 & 0.738 \\
Hit@1 & 0.553 & 0.536 & 0.636 & 0.614 & 0.724 & 0.698 & 0.665 & 0.647 & 0.596 & 0.593 \\
MR    & 44.58 & 45.83 & 60.61 & 63.32 & 60.55 & 66.19 & 39.78 & 41.69 & 60.48 & 65.48 \\
MRR   & 0.676 & 0.664 & 0.732 & 0.716 & 0.789 & 0.769 & 0.752 & 0.736 & 0.682 & 0.679 \\
\bottomrule
\end{tabular}
\normalsize
\end{center}
\end{table*}

\begin{table*}[t]
\begin{center}
\setlength{\tabcolsep}{5pt}
\footnotesize
\caption{The overall performance of \dglke after various 
optimizations with 8 GPUs on Freebase.}
\label{table:overall_opt_acc_freebase}
\begin{tabular}{lcccccccccc}
\toprule
     & \multicolumn{2}{c}{TransE\_l2} & \multicolumn{2}{c}{DistMult} & \multicolumn{2}{c}{ComplEx} &
\multicolumn{2}{c}{RotatE} & \multicolumn{2}{c}{TransR} \\
& 1GPU & Fastest & 1GPU & Fastest & 1GPU & Fastest & 1GPU & Fastest & 1GPU & Fastest \\
\midrule
Hit@10 & 0.865 & 0.822 & 0.839 & 0.837 & 0.837 & 0.830 & 0.750 & 0.730 & N/A & 0.765 \\
Hit@3 & 0.823 & 0.759 & 0.813 & 0.810 & 0.812 & 0.803 & 0.718 & 0.697 & N/A & 0.723 \\
Hit@1 & 0.771 & 0.669 & 0.785 & 0.780 & 0.785 & 0.773 & 0.668 & 0.653 & N/A & 0.545 \\
MR    & 31.64 & 38.44 & 44.93 & 48.58 & 47.79 & 51.40 & 187.7 & 197.51 & N/A & 103.06 \\
MRR   & 0.806 & 0.726 & 0.805 & 0.801 & 0.804  & 0.794 & 0.699 & 0.682 & N/A & 0.642 \\
\bottomrule
\end{tabular}
\normalsize
\end{center}
\end{table*}

\begin{figure}
\centering
\includegraphics[scale=0.3]{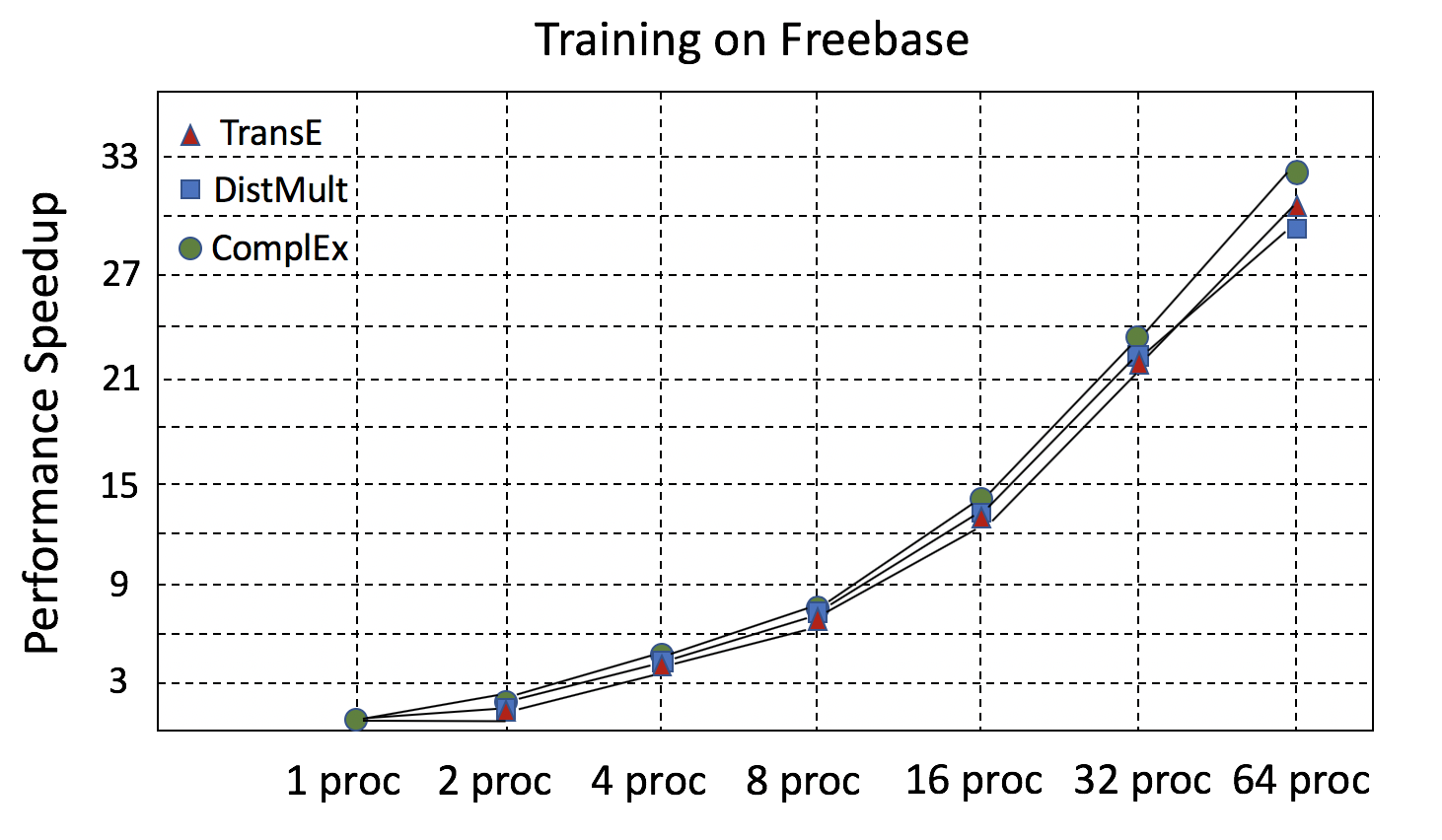}
\caption{Speedup of many-core training.}
\label{fig:cpu-speedup}
\end{figure}

With a maximum speedup of $11\times$ with single-GPU training, we sacrifice 
little on accuracy. Table~\ref{table:overall_opt_acc_fb15k} and 
Table~\ref{table:overall_opt_acc_freebase}
shows the accuracy of \dglke with 1 and 8 GPUs on FB15k and Freebase. 
The \textit{1GPU} columns shows the baseline accuracy and 
the \textit{Fastest} shows the accuracy with the 
fastest configuration on 8 GPUs. For FB15k, we achieve the fastest
training speed with 8 processes on 8 GPUs, while for Freebase, we use 
8 GPUs and 16 concurrent processes. In all experiments,
the total number of epochs we run is the same for both the \textit{1GPU} and \textit{Fastest} settings. Here, we only show TransR with 8 GPUs
on Freebase because training TransR on one GPU takes very long time.

\subsection{Many-core training} \label{eval-many-core}
Many of the techniques illustrated in multi-GPU training can also
be applied to multi-core training.
Figure~\ref{fig:cpu-speedup} shows that \dglke scales well on
an r5dn instance with 48 CPU cores. The training accuracy result of 
TransE and DistMult with 48 CPU cores is shown in Table~\ref{table:acc_wn18}
(column labeled ``Single'').

\subsection{Distributed training} \label{eval-dist}
\begin{figure}
\centering
\includegraphics[scale=0.42]{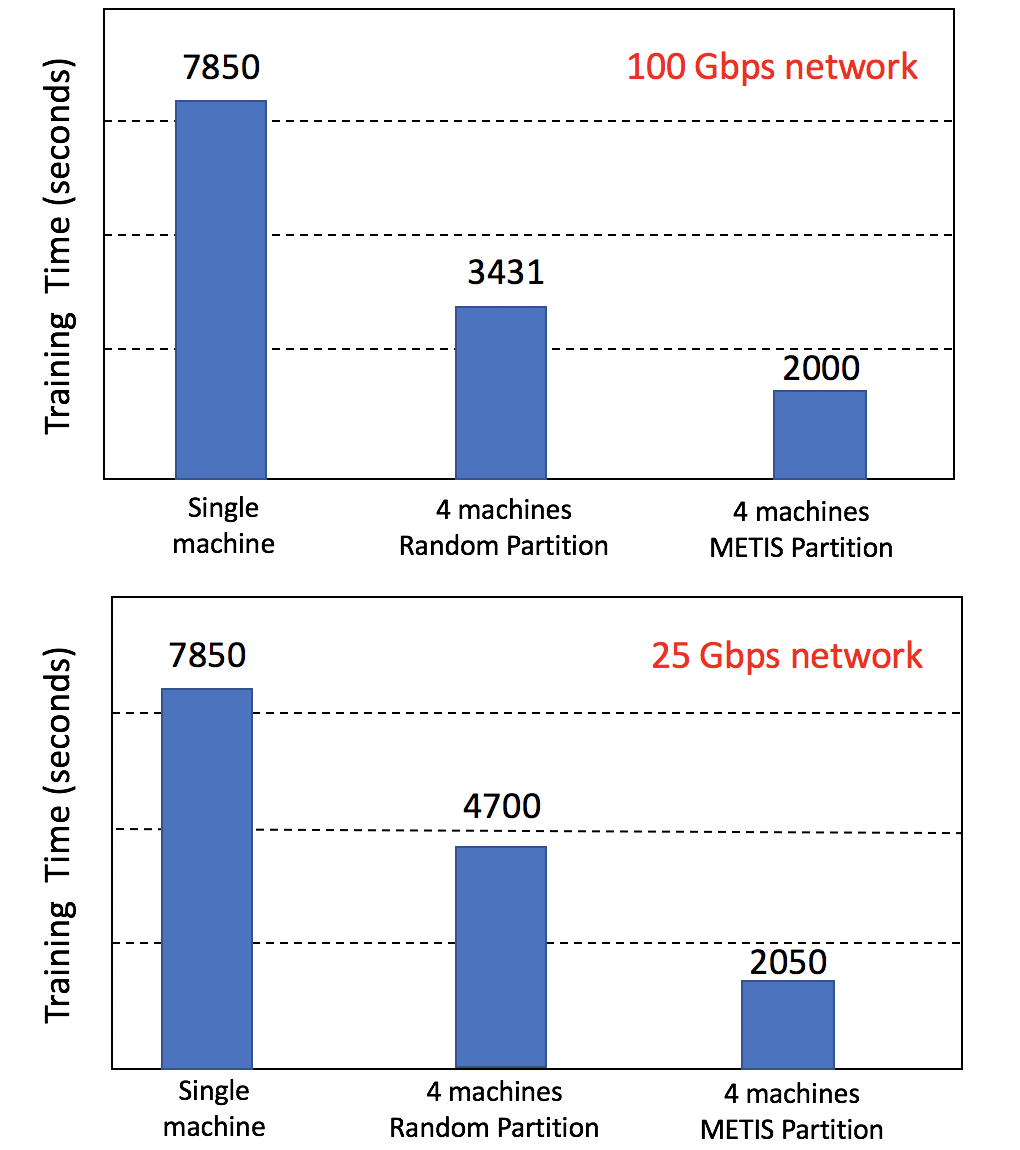}
\caption{The runtime of distributed training.}
\label{fig:dist}
\end{figure}

In distributed training, we use 4 r5dn.24xlarge EC2 instances as our cluster environment. In this section, we compare the baseline single-machine training with distributed training using both random partitioning and METIS partitioning on Freebase.

METIS partitioning on distributed training gets nearly $3.5\times$ speedup compared with the single-machine baseline (Figure~\ref{fig:dist}) without sacrificing any model accuracy (Table~\ref{table:acc_wn18}). The training speed of using METIS partitioning get about 20\% speedup over random partitioning because METIS partitioning leads to much lower overhead than random partitioning.

\begin{table}
\begin{center}
\setlength{\tabcolsep}{5pt}
\footnotesize
\caption{The accuracy of random graph partition and METIS graph partition for distributed training.}
\label{table:acc_wn18}
\begin{tabular}{lcccccc}
\toprule
       & \multicolumn{3}{c}{TransE} & \multicolumn{3}{c}{DistMult} \\
       & Single & Random & METIS & Single & Random & METIS              \\
\midrule
Hit@10& 0.796 & 0.790 & 0.790 & 0.751 & 0.739 & 0.731              \\
Hit@3 & 0.734 & 0.735 & 0.726 & 0.712 & 0.709 & 0.700              \\
Hit@1 & 0.634 & 0.689 & 0.634 & 0.696 & 0.619 & 0.612           \\
MR    & 54.51 & 64.05 & 34.59 & 123.1 & 128.23 & 136.19          \\
MRR   & 0.696 & 0.726 & 0.692 & 0.68 & 0.692 & 0.691              \\
\bottomrule
\end{tabular}
\normalsize
\end{center}
\end{table}

\subsection{Overall performance} \label{eval-compare}

We evaluate \dglke on the datasets in Table \ref{datasets} and compare with
two existing packages: GraphVite and PBG on both CPUs and GPUs.
Because GraphVite and PBG only provide a subset of the models
in \dglke, we only compare with them with the models available in these two
packages. 



\begin{figure}
\centering
\includegraphics[scale=0.3]{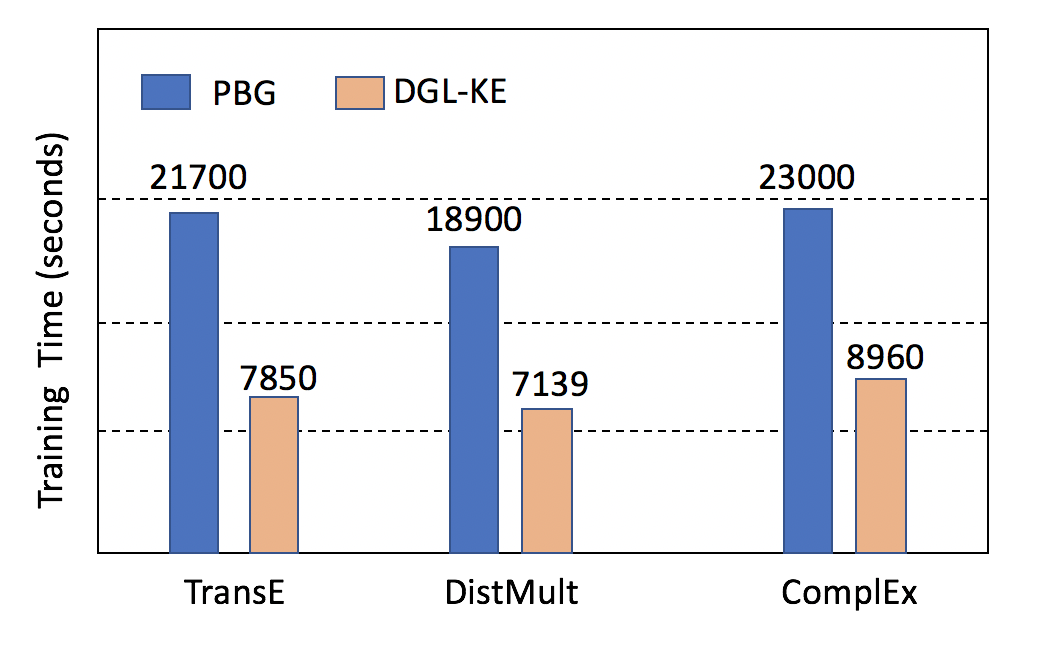}
\caption{The runtime of PBG and \dglke on Freebase.}
\label{fig:vs-pbg}
\end{figure}

\begin{figure}
\centering
\includegraphics[scale=0.3]{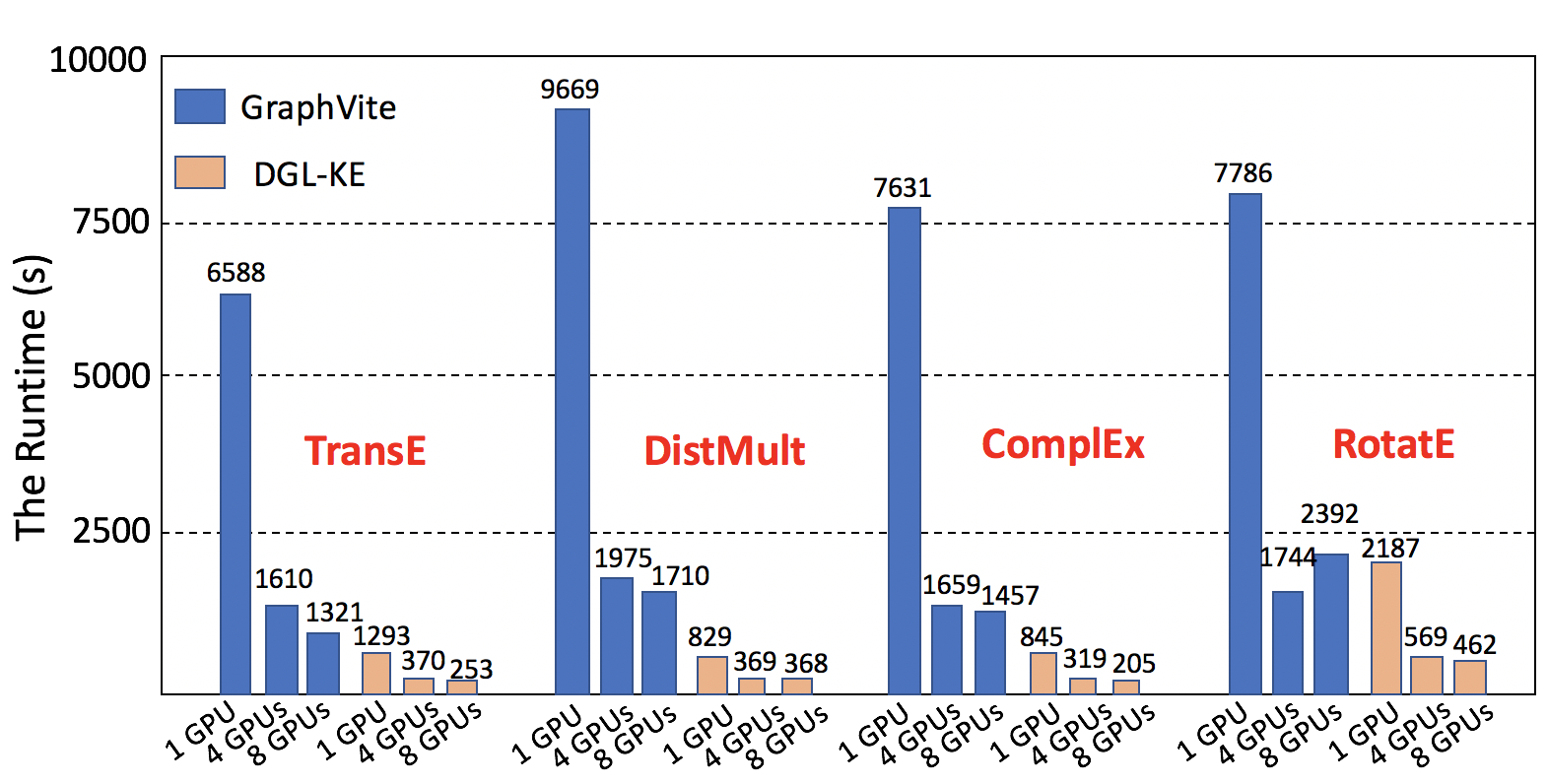}
\caption{The runtime of GraphVate and \dglke on FB15k.}
\label{fig:vs-graphvite}
\end{figure}

\begin{figure}
\centering
\includegraphics[scale=0.3]{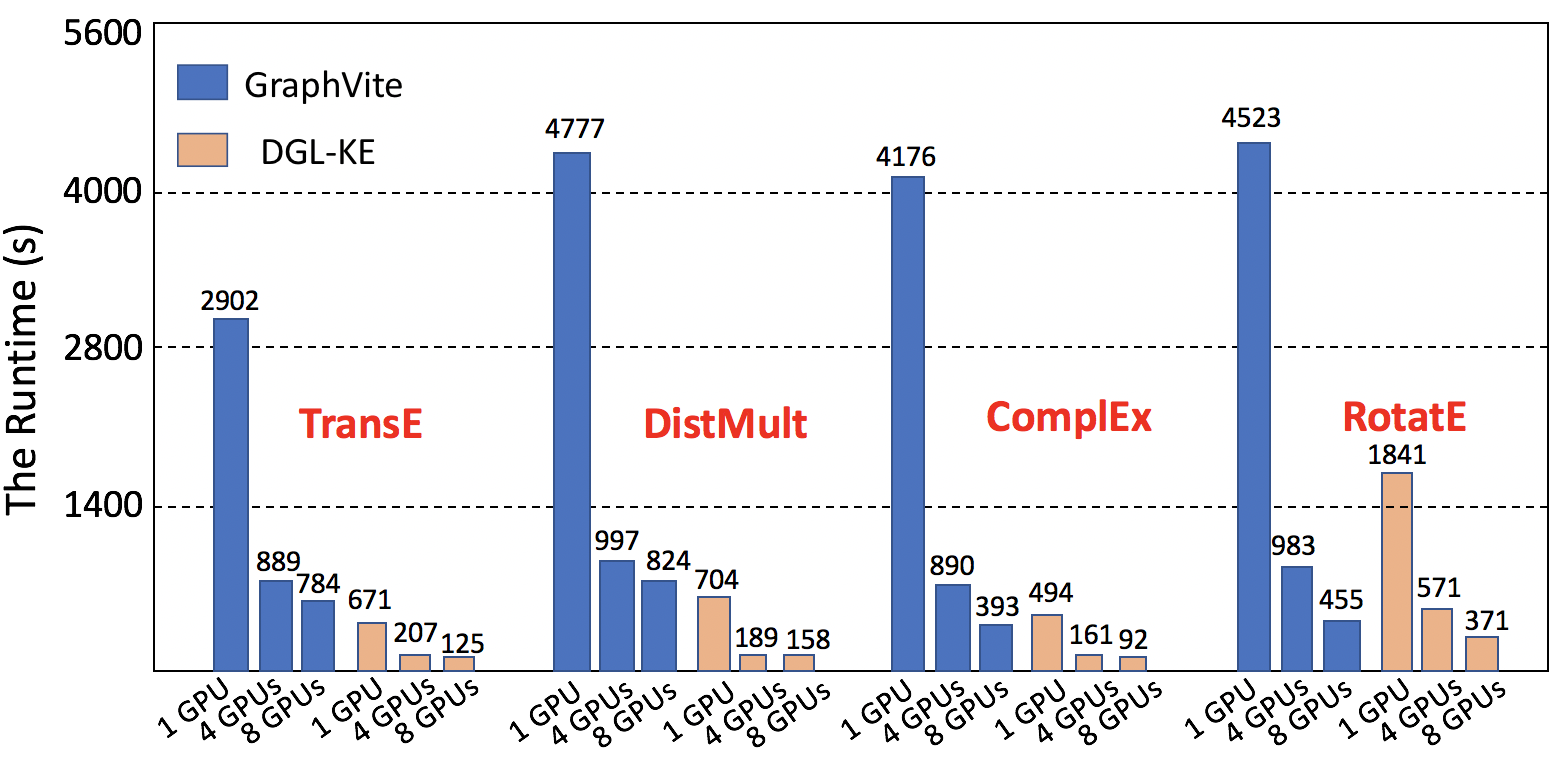}
\caption{The runtime of GraphVate and \dglke on WN18.}
\label{fig:vs-graphvite-wn}
\end{figure}

\subsubsection{Comparison with GraphVite}

\dglke is consistently faster than  GraphVite on both FB15k and WN18 
(Figure~\ref{fig:vs-graphvite} and Figure~\ref{fig:vs-graphvite-wn}) when  training 
all KGE models to reach similar accuracy (Table~\ref{table:acc_fb15k_graphvite} and
Table~\ref{table:acc_wn18_graphvite}).
For most of the models, \dglke is $5 \times$ faster than
GraphVite. This is mainly due to \dglke converges faster 
than GraphVite. In all cases, \dglke only needs less than 100
epochs to converge but GraphVite needs thousands of epochs.
When evaluating GraphVite, we use the recommended configuration by the package
for each algorithm when running on 1 GPU and 4 GPUs, while having
some hyperparameter tuning for 8 GPUs to get compatible results with 1 GPU runs.
When evaluating \dglke, we use the same dimension size of entity and relation embedding as
GraphVite, but tune hyper-parameters
such as learning rate, negative sample size and batch size, for better accuracy.

\begin{table}
\begin{center}
\setlength{\tabcolsep}{5pt}
\footnotesize
\caption{The accurancy of \dglke and GraphVite on FB15k with 1, 4 and 8 GPUs.}
\label{table:acc_fb15k_graphvite}
\begin{tabular}{lcccccc}
\toprule
       & \multicolumn{6}{c}{TransE}  \\
       & \multicolumn{3}{c}{\dglke} & \multicolumn{3}{c}{GraphVite} \\
       & 1 GPU & 4 GPU & 8 GPU & 1 GPU & 4 GPU & 8 GPU              \\
\midrule
Hit@10 & 0.873 & 0.866 & 0.863 & 0.869 & 0.873 & 0.872              \\
Hit@3  & 0.801 & 0.791 & 0.789 & 0.793 & 0.791 & 0.781              \\
Hit@1  & 0.612 & 0.613 & 0.611 & 0.606 & 0.586 &0.373           \\
MR     & 40.84 & 44.52 & 45.12 & 37.81 & 38.89 & 40.63              \\
MRR    & 0.717 & 0.713 & 0.711 & 0.711 & 0.700 & 0.588              \\
\bottomrule
\\
\toprule
       & \multicolumn{6}{c}{DistMult} \\
        & \multicolumn{3}{c}{\dglke} & \multicolumn{3}{c}{GraphVite} \\
       & 1 GPU & 4 GPU & 8 GPU & 1 GPU & 4 GPU & 8 GPU \\
\midrule
Hit@10 & 0.895 & 0.890 & 0.882 & 0.892 & 0.876 & 0.873 \\
Hit@3  & 0.835 & 0.825 & 0.806 & 0.834 & 0.814 & 0.800 \\
Hit@1  & 0.702 & 0.680 & 0.645 & 0.715 & 0.697 & 0.646 \\
MR     & 44.50 & 51.79 & 56.54 & 40.51 & 69.15 & 60.11 \\
MRR    & 0.777 & 0.762 & 0.736 & 0.783 & 0.765 & 0.733 \\
\bottomrule
\\
\toprule
& \multicolumn{6}{c}{ComplEx} \\
& \multicolumn{3}{c}{\dglke} & \multicolumn{3}{c}{GraphVite} \\
       & 1 GPU & 4 GPU & 8 GPU & 1 GPU & 4 GPU & 8 GPU \\
\midrule
Hit@10 & 0.892 & 0.881 & 0.879 & 0.867 & 0.830 & 0.810 \\
Hit@3  & 0.839 & 0.824 & 0.816 & 0.788 & 0.742 & 0.718 \\
Hit@1  & 0.735 & 0.705 & 0.694 & 0.643 & 0.591 & 0.572 \\
MR     & 50.47 & 68.17 & 70.13 & 58.68 & 153.4 & 145.6 \\
MRR    & 0.795 & 0.773 & 0.764 & 0.727 & 0.679 & 0.660 \\
\bottomrule
\\
\toprule
& \multicolumn{6}{c}{RotatE} \\
& \multicolumn{3}{c}{\dglke} & \multicolumn{3}{c}{GraphVite} \\
       & 1 GPU & 4 GPU & 8 GPU & 1 GPU & 4 GPU & 8 GPU \\
\toprule
Hit@10 & 0.888 & 0.883 & 0.881 & 0.875 & 0.892 & 0.887 \\
Hit@3  & 0.820 & 0.813 & 0.812 & 0.814 & 0.830 & 0.823 \\
Hit@1  & 0.647 & 0.640 & 0.648 & 0.691 & 0.688 & 0.646 \\
MR     & 34.38 & 35.47 & 35.71 & 41.75 & 35.87 & 43.26 \\
MRR    & 0.744 & 0.737 & 0.740 & 0.762 & 0.768 & 0.743 \\
\bottomrule
\end{tabular}
\normalsize
\end{center}
\end{table}

\begin{table}
\begin{center}
\setlength{\tabcolsep}{5pt}
\footnotesize
\caption{The accuracy of \dglke and GraphVite on wn18 with 1, 4 and 8 GPUs.}
\label{table:acc_wn18_graphvite}
\begin{tabular}{lcccccc}
\toprule
       & \multicolumn{6}{c}{TransE}  \\
       & \multicolumn{3}{c}{\dglke} & \multicolumn{3}{c}{GraphVite} \\
       & 1 GPU & 4 GPU & 8 GPU & 1 GPU & 4 GPU & 8 GPU              \\
\midrule
Hit@10& 0.950 & 0.950 & 0.948 & 0.953 & 0.943 & 0.950              \\
Hit@3 & 0.930 & 0.928 & 0.926 & 0.888 & 0.905 & 0.910              \\
Hit@1 & 0.600 & 0.592 & 0.528 & 0.582 & 0.577 &0.346           \\
MR    & 343.5 & 355.7 & 329.6 & 260.4 & 394.3 & 342.9          \\
MRR   & 0.763 & 0.759 & 0.726 & 0.739 & 0.741 & 0.627              \\
\bottomrule\\\toprule
       & \multicolumn{6}{c}{DistMult} \\
       & \multicolumn{3}{c}{\dglke} & \multicolumn{3}{c}{GraphVite} \\
      & 1 GPU & 4 GPU & 8 GPU & 1 GPU & 4 GPU & 8 GPU \\
\midrule
Hit@10& 0.945 & 0.936 & 0.939 & 0.955 & 0.944 & 0.937 \\
Hit@3 & 0.918 & 0.910 & 0.912 & 0.922 & 0.917 & 0.897 \\
Hit@1 & 0.702 & 0.687 & 0.659 & 0.710 & 0.715 & 0.645 \\
MR    & 587.7 & 725.3 & 637.7 & 313.6 & 598.4 & 657.2 \\
MRR   & 0.812 & 0.800 & 0.786 & 0.819 & 0.818 & 0.772 \\
\bottomrule\\\toprule
& \multicolumn{6}{c}{ComplEx} \\
& \multicolumn{3}{c}{\dglke} & \multicolumn{3}{c}{GraphVite} \\
      & 1 GPU & 4 GPU & 8 GPU & 1 GPU & 4 GPU & 8 GPU \\
\midrule
Hit@10& 0.961 & 0.954 & 0.950 & 0.894 & 0.903 & 0.914 \\
Hit@3 & 0.950 & 0.948 & 0.944 & 0.872 & 0.844 & 0.895 \\
Hit@1 & 0.918 & 0.940 & 0.938 & 0.829 & 0.882 & 0.860 \\
MR    & 259.9 & 599.9 & 828.3 & 1218.8 & 1255.7 & 1273.7 \\
MRR   & 0.935 & 0.945 & 0.942 & 0.854 & 0.866 & 0.881 \\
\bottomrule\\\toprule
& \multicolumn{6}{c}{RotatE} \\
& \multicolumn{3}{c}{\dglke} & \multicolumn{3}{c}{GraphVite} \\
      & 1 GPU & 4 GPU & 8 GPU & 1 GPU & 4 GPU & 8 GPU \\
\midrule
Hit@10& 0.955 & 0.955 & 0.954 & 0.960 & 0.957 & 0.956 \\
Hit@3 & 0.949 & 0.948 & 0.948 & 0.953 & 0.950 & 0.950 \\
Hit@1 & 0.942 & 0.942 & 0.942 & 0.944 & 0.942 & 0.941 \\
MR    & 390.6 & 379.9 & 384.296 & 231.1 & 311.1 & 367.5 \\
MRR   & 0.946 & 0.946 & 0.946 & 0.949 & 0.947 & 0.947 \\
\bottomrule
\end{tabular}
\normalsize
\end{center}
\end{table}


\subsubsection{Comparison with PBG} 

\dglke runs twice as fast as PBG when training KGE models on Freebase (Figure \ref{fig:vs-pbg}). There are 
many factors that contribute to the slower training speed in PBG. One of the major 
factors is that PBG handles relation embeddings as dense model weights. As such, 
the computation in a batch involves in all relation embeddings in the graph, 
which is 10 times more than necessary on Freebase. In contrast, \dglke reduces 
the number of relation embeddings involved in a batch and significantly reduces 
the amount of computation and data movement.

\section{Conclusions}
We develop an efficient package called \dglke to train knowledge graph embeddings at large scale. It implements a number of optimization techniques to improve locality, reduce data communication, while harnessing parallel computing capacity. As a result, \dglke significantly 
outperforms the state-of-the-art packages for knowledge graph embeddings on a variety of hardware, including many-core CPU, multi-GPU as well as cluster of machines.
Our experiments show that \dglke scales well with machine resources
almost linearly while still achieving very high model accuracy. \dglke is available at https://github.com/awslabs/dgl-ke.


\section{Acknowledgments}
We thank the RotatE authors for making their knowledge graph embedding package
KnowledgeGraphEmbedding open-source. \dglke was built based on their package.

\bibliographystyle{named}
\bibliography{dglke}

\end{document}